\documentclass[5p]{elsarticle}
\usepackage{amssymb,amsmath}

\usepackage{epsfig}
\usepackage{graphics}

\newcommand{\vol}{\mathop{\rm vol}\nolimits}

\newcommand{\pa}{\partial}
\newcommand{\ii}{{\rm i}}
\newcommand{\ee}{{\rm e}}

\newcommand{\ve}{\varepsilon}
\newcommand{\ZZ}{\mathbb Z}
\newcommand{\RR}{\mathbb R}
\newcommand{\half}{\textstyle\frac{1}{2}}

\begin{document}

\journal{Physics Letters B}

\begin{frontmatter}

\title{Generalized Skyrme Crystals}

\author[dur]{J.\ Silva Lobo\corref{cor1}}
\ead{j.i.silva-lobo@durham.ac.uk}

\author[dur]{R.\ S.\ Ward}  
\ead{richard.ward@durham.ac.uk}

\cortext[cor1]{Corresponding author}

\address[dur]{Department of Mathematical Sciences, Durham University, South Road, Durham, DH1 3LE, United Kingdom}

\begin{abstract}
This letter deals with triply-periodic (crystalline) solutions
in a family of Skyrme systems, namely where the field takes values in the
squashed 3-sphere. The family includes the standard Skyrme model
(round 3-sphere), and the Skyrme-Faddeev case (maximal squashing).
In the round case, the lowest-energy crystal is the well-known cubic
lattice of half-skyrmions; but in the squashed case the minimal-energy
crystal structures turn out to be different. We describe some of the
solutions that arise, including arrays of vortices and multi-sheeted
structures.
\end{abstract}


\end{frontmatter}




\section{Introduction}

In the basic SU(2) Skyrme model, the solution with the lowest energy-per-charge
$E_N$ is the triply-periodic skyrme crystal \cite{MS04}. For a given value of
the topological charge (winding number, or baryon number) $N$, there are in
general many isolated $N$-skyrmion solutions, {\sl i.e.}\ local
minima of the energy; but the belief is that for $N\gg1$, the
lowest-energy solution will resemble a chunk of the skyrme crystal.

However, in the Skyrme-Faddeev system, where the field takes values in $S^2$
and the topological charge $N$ is the Hopf number, the situation is less clear.
If there were a triply-periodic crystalline solution with $E_N=c$
(where $c$ is a constant independent of $N$), as in the Skyrme case,
then a large crystalline chunk would have $E_N\sim c$.
But we know \cite{Y04} that there exist solutions with
$E_N\sim N^{-1/4}$, so such a crystal chunk certainly could not be the global
energy minimum among fields of charge $N$, although it could be a local
minimum. In general, it seems to be the case that Hopf solitons tend to clump
into a tangle; various examples, for low values of $N$, may be seen in
\cite{BS99, HS99, HJS04, S07, JH09}.

The aim of this letter is to explore this difference in behaviour, by
investigating triply-periodic crystalline solutions in a system which
interpolates between the two cases above. This generalized Skyrme system
was introduced in \cite{W04}; it is labelled by a parameter $\alpha\in[0,1]$,
in such a way that $\alpha=0$ gives the `pure' Skyrme model, whereas
$\alpha=1$ gives the Skyrme-Faddeev model. In particular, we wish to see
what happens to the skyrme crystal as the system deforms away from the basic
Skyrme case.

The interpolating system is very natural geometrically. For simplicity,
let us restrict to static fields in all of what follows. Let $\Sigma$ be a
compact 3-dimensional, oriented, connected Riemannian manifold.
Then for fields $\Phi:\RR^3\to\Sigma$, the Skyrme energy functional
$E[\Phi]$ has a natural definition, depending in particular on the metric
of $\Sigma$ \cite{M87}.  With the usual boundary condition
$\Phi(x^j)\to \Phi_0$ as $|x^j|\to \infty$, where $\Phi_0$ is some specified
point on $\Sigma$, the topological charge $N$ is defined to be
the degree of the map $\Phi$.
Our system is obtained by taking $\Sigma$ to be the squashed 3-sphere
(Berger sphere): the standard 3-sphere squashed along its Hopf
fibres by a factor $1-\alpha$. If $\alpha=0$, there is no squashing:
the target space is just the round sphere SU(2), and we have the Skyrme model.
In the degenerate case $\alpha=1$, the 3-sphere becomes a 2-sphere, the
winding number $N$ becomes the Hopf number, and we have the Skyrme-Faddeev
model.

The system has a potential condensed-matter interpretation in which the
two complex fields $(Z_1,Z_2)$ represent two flavours of Cooper pairs
\cite{BFN02}. The parameter $\alpha$ then appears as the coefficient of a
term $J_{\mu} J^{\mu}$, where $J_{\mu}=\ii Z^\dagger\,\pa_{\mu}Z$ is the current
density. In particular, arrays might be of interest in this context.


\section{The Generalized Skyrme System}

In this section, we give some details of the one-parameter interpolating
family of Skyrme systems. The field is denoted
$\Phi_{\beta}=(\Phi_1,\Phi_2,\Phi_3,\Phi_4)$, where
each component $\Phi_{\beta}$ is a function of the spatial coordinates
$x^j=(x^1,x^2,x^3)=(x,y,z)$, and the constraint $\Phi_{\beta}\Phi_{\beta}=1$
is imposed. The energy density for the system, parametrized by
$\alpha\in[0,1]$, is
\begin{equation} \label{Enden}
  {\cal E}=\lambda_2[(\pa_j\Phi_{\beta})(\pa_j\Phi_{\beta})-\alpha P_jP_j]
  +\lambda_4[2(1-\alpha)F^j_{\beta\gamma}F^j_{\beta\gamma}+\alpha Q^jQ^j],
\end{equation}
where
\begin{eqnarray*}
P_j &=& \Omega_{\beta\gamma}\Phi_{\beta}\pa_j\Phi_{\gamma},  \\
F^j_{\beta\gamma} &=&
    \half\ve^{jkl}(\pa_k\Phi_{\beta})(\pa_l\Phi_{\gamma}),  \\
Q^j &=& \Omega_{\beta\gamma}F^j_{\beta\gamma},
\end{eqnarray*}
with $\Omega_{\beta\gamma}$ being a symplectic form with non-zero
components $\Omega_{12}=-\Omega_{21}=-\Omega_{34}=\Omega_{43}=1$.
The coupling constants $\lambda_2$ and $\lambda_4$
can be scaled as desired, by changing the units of energy and length;
in what follows, we shall use the `geometrical' choice
$\lambda_2=1/[4\pi^2(3-\alpha)]$ and $\lambda_4=1/[4\pi^2(3-2\alpha)]$
as in \cite{W04}. The effect of this is that the energy of a 1-skyrmion
on $\RR^3$ is approximately independent of $\alpha$ ($E_1\approx1.22$).

If we are thinking of the Skyrme model as a nonlinear theory of pions,
then the deformation when $\alpha$ becomes positive may be viewed as
putting one of the three
pions on a different footing from the other two. To see this explicitly,
we impose the boundary condition $\Phi_{\mu}\to(0,0,0,1)$ as $|x^j|\to \infty$,
and identify $(\Phi_1,\Phi_2,\Phi_3)$ as the pion fields in the usual way.
Then in the asymptotic region the energy density (\ref{Enden}) becomes
\begin{equation} \label{Enden_asymp}
  {\cal E}\approx\lambda_2[(\pa_j\Phi_1)^2+(\pa_j\Phi_2)^2+(1-\alpha)(\pa_j\Phi_3)^2].
\end{equation}

If $\alpha=1$, then one of the fields becomes non-dynamic. This amounts to
passing from the original target space $S^3\cong{\rm SU(2)}$ to the quotient space
$S^2\cong{\rm SU(2)/U(1)}$, by the Hopf projection
\begin{eqnarray}
\psi_1 &=& 2(\Phi_1\Phi_3-\Phi_2\Phi_4),   \nonumber \\
\psi_2 &=& 2(\Phi_2\Phi_3+\Phi_1\Phi_4),   \nonumber \\
\psi_3 &=& \Phi_3^4+\Phi_4^2-\Phi_1^2-\Phi_2^2. \label{hopfproj}
\end{eqnarray}
The relevant field is then the unit 3-vector $\vec\psi$, and the
energy density (\ref{Enden}) becomes
\begin{equation} \label{Enden_Hopf}
  {\cal E} = \frac{1}{32\pi^2}\left[(\pa_j\vec\psi)^2
       + \frac{1}{4}(G_{jk})^2\right],
\end{equation}
where $G_{jk}=
\vec\psi\cdot(\pa_j\vec\psi)\times(\pa_k\vec\psi)$. This is
the Skyrme-Faddeev energy \cite{FN97}.

The energy $E[\Phi]$ is the integral of (\ref{Enden}) over $\RR^3$,
or over a fundamental cell if one is dealing with periodic fields.
There is a Bogomolny-type lower bound \cite{M87} on the energy of fields
with topological charge $N$, namely
$E\geq6N\sqrt{\lambda_2 \lambda_4}\vol{(\Sigma)}$.
Using the fact that the volume of the squashed 3-sphere is
$2\pi^2\sqrt{1-\alpha}$, and with our choice of $\lambda_2$ and $\lambda_4$,
this gives
\begin{equation} \label{Bog}
    E_N \geq \frac{\sqrt{1-\alpha}}{\sqrt{(1-\alpha/3)(1-2\alpha/3)}}.
\end{equation}

In the limit $\alpha=1$, the bound (\ref{Bog}) becomes trivial.
There are various nontrivial bounds in this case.
For isolated Hopf solitons on $\RR^3$, we know that $E_N\geq cN^{-1/4}$,
where $c$ is a constant; it is conjectured that this bound holds for
$c=1$ \cite{W99}. This bound is not valid for triply-periodic fields
$\Phi:T^3\to S^2$; but in that case
there is another lower bound, which follows from the general formula
given in \cite{Sp10}: if $\Phi$ is triply-periodic, and $L$ is the largest
of its periods, then its normalized energy satisfies $E_N\geq\pi/(2L)$.

For $\alpha=0$, the expression (\ref{Enden}) gives the standard static
Skyrme energy, and it has an obvious O(4) symmetry. For $\alpha\neq0$,
the symmetry is broken to U(2), this being the subgroup of O(4) which
preserves the symplectic form $\Omega$.

In this letter, we are focusing on crystalline configurations, so
the fields are periodic in $(x^1,x^2,x^3)$ with periods $(L_1,L_2,L_3)$
respectively. In effect, $\Phi$ is a map from $T^3$ to $S^3$, and it
is classified topologically by its degree $N\in\ZZ$. The topological
classification of maps from $T^3$ to $S^2$, which is relevant to the
$\alpha=1$ case, is more complicated. But if we restrict to fields which
are algebraically inessential, meaning that the 2-form $G_{jk}$ belongs to
the trivial cohomology class in $H^2(T^3,\ZZ)$, then the classification
remains a single integer $N\in\ZZ$ ({\sl cf.}\ \cite{KS06}). The
Hopf charge of the Hopf projection $\vec\psi$ of $\Phi$ is equal to the
degree of $\Phi$, so it is consistent to let $N$ denote either.


\section{Crystals in the Generalized Skyrme Family}

In this section, we investigate various triply-periodic solutions
in the family of generalized Skyrme systems para- metrized by $\alpha$.
There are many of these, corresponding to different local minima
of the energy, and we shall describe only some of them.

The results that follow are obtained by numerical minimization of the
energy functional $E[\Phi]$. More specifically, the expression for the
energy is replaced by a second-order finite-difference approximation
on a rectangular lattice with periodic boundary conditions, and the
minima are found by conjugate-gradient relaxation. Testing with
various different values of the lattice spacing indicates that the
accuracy, for example in the energy $E$, is better than $1\%$.
As remarked above, there are many local minima, and the initial
configuration determines which one of these is obtained after flowing
down the energy gradient. We also vary the side-lengths $L_1$, $L_2$
and $L_3$ of the fundamental cell, so as to get the lowest possible
value of $E[\Phi]$. 

In the original Skyrme system ($\alpha=0$), the lowest-energy crystal
is a cubic lattice \cite{MS04}. Its unit cell contains eight half-skyrmions,
and has $N=4$. We begin by taking this as the initial configuration, and
minimizing $E[\Phi]$ for various values of $\alpha$.

In the $\alpha=0$ case, the minimum occurs at $L_1=L_2=L_3=4.7$;
but the side-lengths cease to be all equal when $\alpha>0$. This
occurs because the system has less symmetry than when $\alpha=0$;
in particular, one of the spatial directions, which we take to be the
$x^3$-direction, is now on a different footing from the other two.
For $\alpha>0$, the half-skyrmions become elongated in the $x^3$-direction,
and the optimal side-lengths satisfy $L_1=L_2>L_3$. These aspects are illustrated
in Figure~1. The upper-left plot shows the optimal (minimal-energy)
values of $L_1=L_2$ and $L_3$, as functions of $\alpha$, in the range
$0\leq\alpha\leq0.9$.
\begin{figure}[hbt] 
  \includegraphics[scale=0.65]{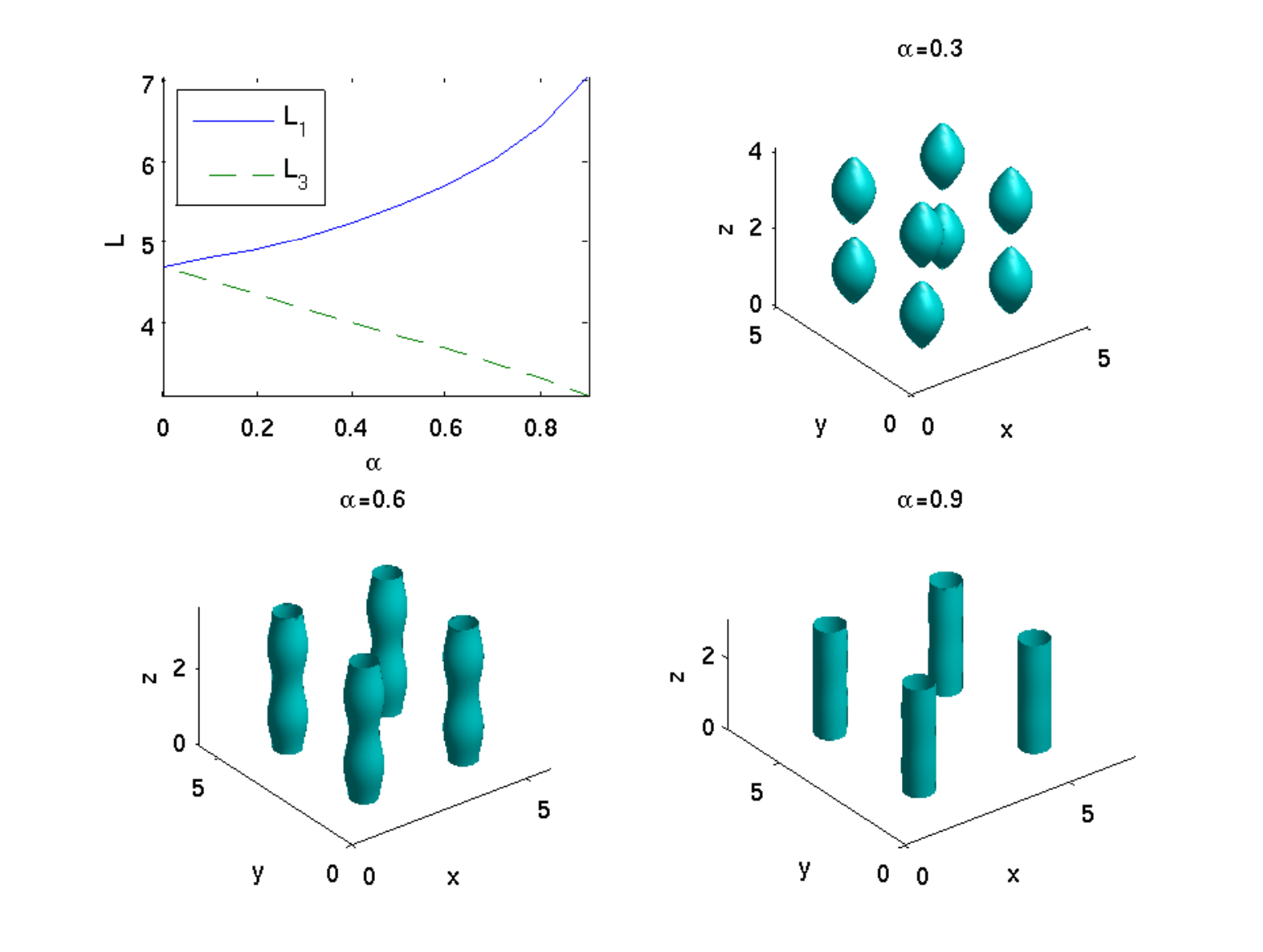}
  \caption{The V+AV+V+AV solution. Optimal values of $L_1=L_2$, and $L_3$;
   and the energy density for $\alpha=0.3,\, 0.6,\, 0.9$.\label{fig1}}
\end{figure}
We see that in the Skyrme case $\alpha=0$, we have $L_1=L_2=L_3=4.7$, as noted
above; but as $\alpha$ increases, the lattice cells are compressed in the
$x^3$-direction and expanded in the $x^1$- and $x^2$-directions. The behaviour
in the $\alpha=1$ limit will be discussed in more detail below.
The other three pictures in Figure~1 plot the energy density ${\cal E}$,
or more precisely the surfaces where ${\cal E}$ equals 0.7 times its
maximum value, for three values of $\alpha$ ranging from 0.3 to 0.9. 
We see that as $\alpha$ increases, the half-skyrmions join up pairwise in the
$x^3$-direction, to form four parallel vortices; these are in fact two vortices
and two antivortices, and so this field is called the V+AV+V+AV solution.
In the $\alpha\to1$ limit, the vortices become homogeneous in the
$x^3$-direction; such vortices will be discussed in more detail below.

The symmetries of the $\alpha>0$ fields depicted in Figure~1 are a
subset of the $\alpha=0$ skyrme-crystal symmetries \cite{MS04}, and
are generated by
\begin{equation} \label{Sym1a}
x^j\mapsto L_j-x^j,\quad \Phi_4\mapsto-\Phi_4\quad{\rm for}\quad j=1,2,3;
\end {equation}
\begin{equation} \label{Sym1b}
x^j\mapsto  L_j/2-x^j,\quad \Phi_j\mapsto-\Phi_j\quad{\rm for}\quad j=1,2,3;
\end {equation}
\begin{eqnarray}
(x^1,x^2,x^3)&\mapsto&(x^2,-x^1,x^3), \nonumber \\
   (\Phi_1,\Phi_2,\Phi_3,\Phi_4)&\mapsto&(\Phi_2,\Phi_1,\Phi_3,-\Phi_4). \label{Sym1c} 
\end {eqnarray}

\bigskip

The normalized energy of these solutions is plotted in Figure~2, as a function
of $\alpha$. 
\begin{figure}[hbt] 
  \includegraphics[scale=0.5]{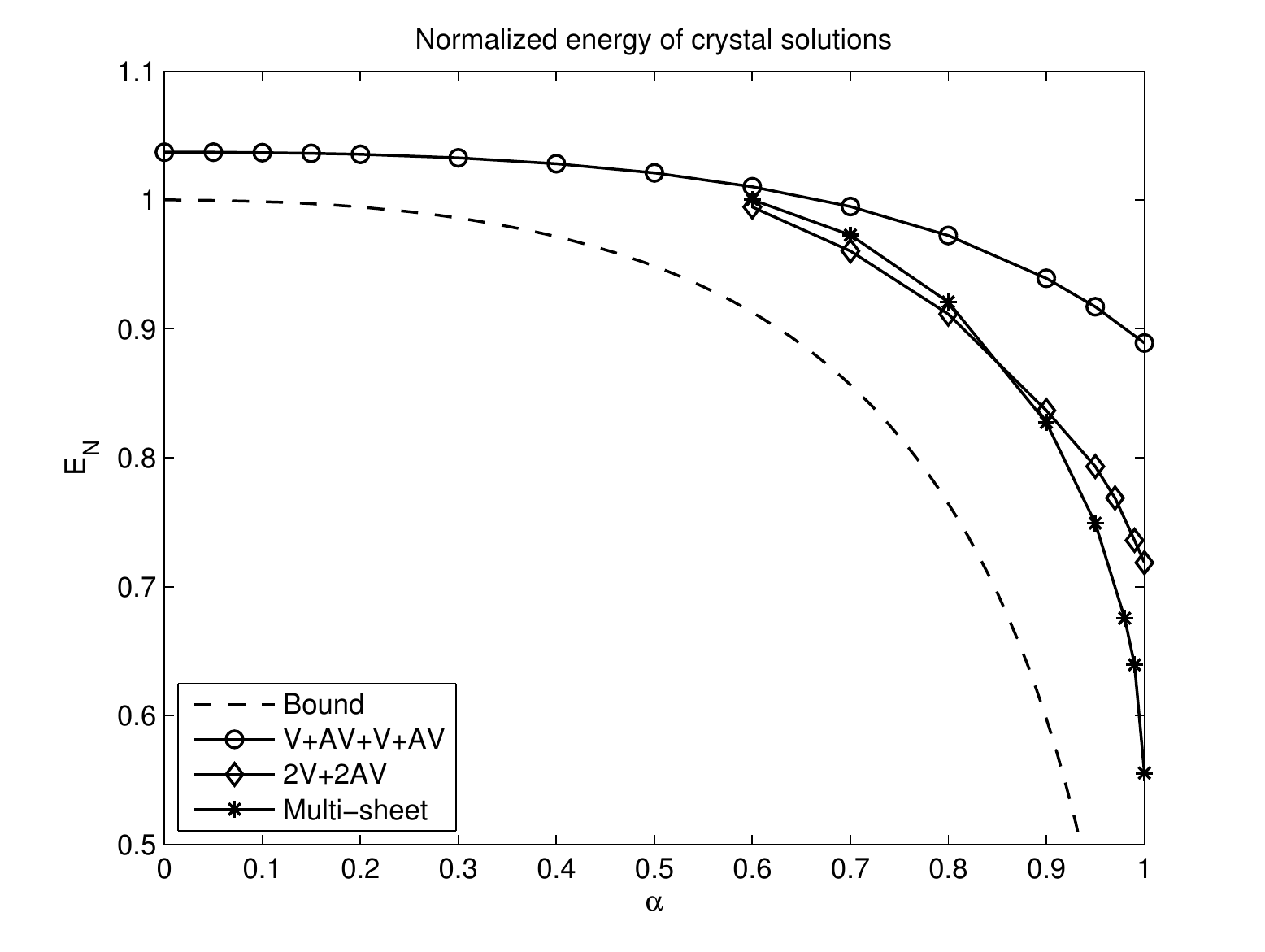}
  \caption{Normalized energy $E_N$ of crystalline solutions,
   versus $\alpha$. \label{fig2}}
\end{figure}
The other plots in Figure~2 are the Bogomolny bound (\ref{Bog}), and the
energies of two other triply-periodic solutions which will be described below.

At this point, let us say more about $x^3$-homogeneous vortices. 
The field of a $p$-vortex with unit
$x^3$-twist, located on the $x^3$-axis, can be put in the form
\begin{equation} \label{Phi-vortex}
  \Phi_1+\ii\Phi_2\approx(x^1+\ii x^2)^p, \quad 
       \Phi_3+\ii\Phi_4\approx\exp{(2\pi\ii x^3/L_3)}
\end {equation}
near that axis. If $p>0$, we refer to it as a $p$-vortex, whereas if $p<0$ we
refer to it as a $p$-antivortex. If we restrict the Hopf projection $\psi$ of
$\Phi$ to the
plane $x^3=0$, we get a map from $T^2$ to $S^2$; the degree of this map is
the total vortex number in the $x^3$-direction (taking the multiplicity $p$ of
each vortex into account). This total vortex number has to be zero, since it
is the degree of a Hopf projection. In other words, there have to be an equal
number of vortices and antivortices.
The topological charge $N$ is the sum of the absolute values of the vortex
numbers; in other words, vortices and antivortices both contribute positively
to $N$. The $\alpha=0.9$ field illustrated in Figure~1 is close
to an $N=4$ multi-vortex configuration, with two 1-vortices and two
1-antivortices, and hence is referred to as V+AV+V+AV.

It is worth noting that analogous V+AV+V+AV configurations can appear in the
$\alpha=0$ case, for values of the side-lengths $L_j$ which differ from the
optimal ones \cite{L10}.

Skyrme vortices, or more precisely vortex-antivortex pairs, were discussed
in \cite{HW08}. In particular, it was shown there that if one has a
parallel vortex-antivortex pair separated by a (large) distance $D$,
then there is an attractive force between them, the leading term of which
is proportional to $1/D$. In our $\alpha$-family, the expression for the
attractive force is the same, except that it acquires a factor of $1-\alpha$.
In other words, vortices and antivortices attract one another, as long as
$\alpha<1$. In the $\alpha\to1$ limit, however, the leading-order term
vanishes. It is not known whether there is still an attractive vortex-antivortex
force in this limit, but numerical experiments suggest that the force becomes
repulsive. In particular, the two vortices and two antivortices in our picture
repel each other when $\alpha=1$, and therefore $L_1,L_2\to\infty$
as $\alpha\to1$, consistent with the upper-left plot in Figure~1.

The second vortex-type solution featuring in Figure~2 is
2V+2AV, and it consists of a 2-vortex and a parallel 2-antivortex. Its energy
is plotted for $\alpha\geq0.6$, and it is apparent that in this range,
its energy is less than that of V+AV+V+AV. In this case, the numerical
evidence again suggests that $L_1,L_2\to\infty$ as $\alpha\to1$, implying
that the 2-vortex and the 2-antivortex repel each other in that limit. The four
pictures in Figure~3 plot the energy density isosurfaces where ${\cal E}$ equals
0.8 times its maximum value, for four values of $\alpha$ ranging from 0.6 to 0.99.
\begin{figure}[hbt] 
  \includegraphics[scale=0.45]{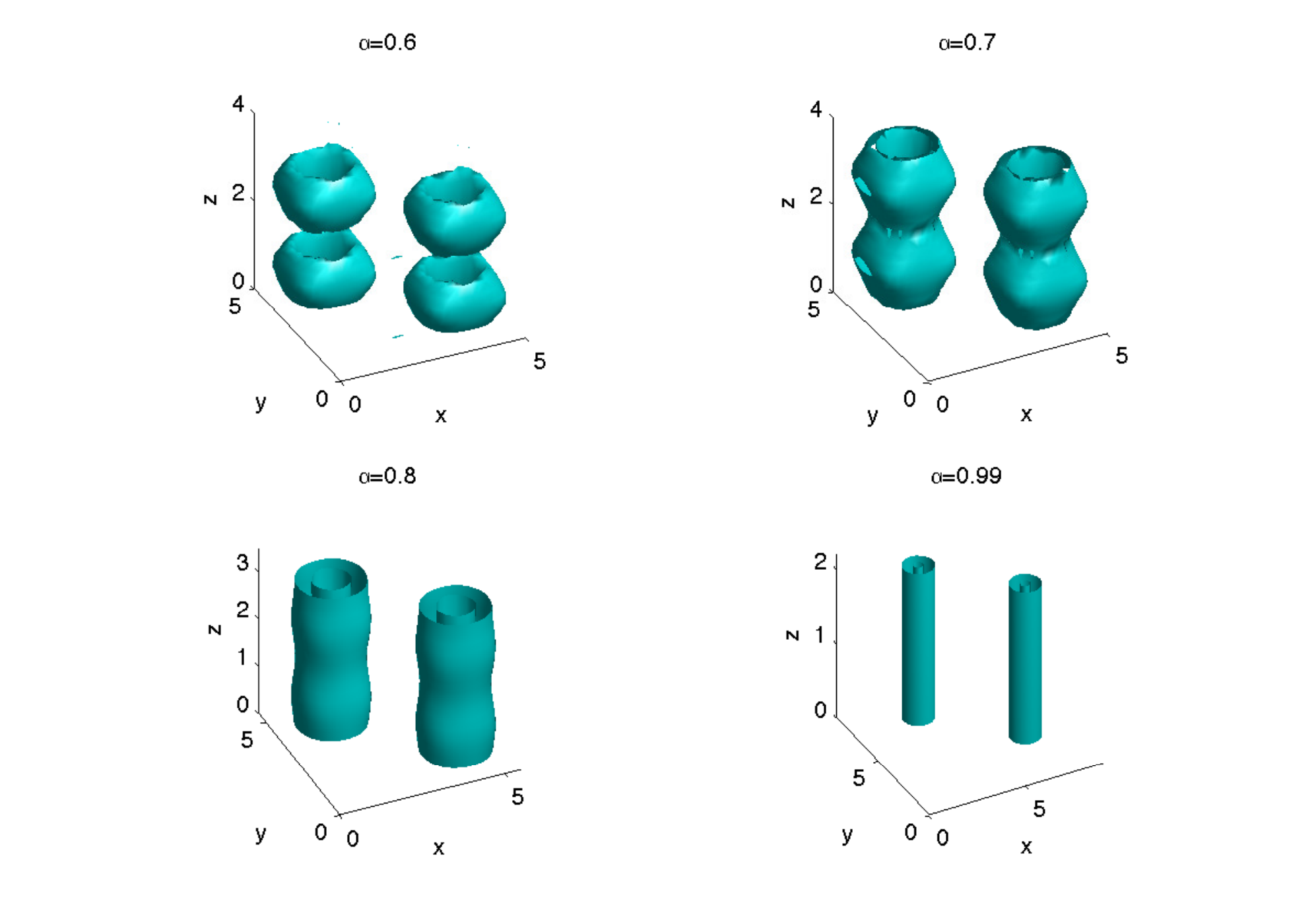}
  \caption{The 2V+2AV solution: energy densities for $\alpha=0.6,\, 0.7,\, 0.8,\,0.99$.\label{fig3}}
\end{figure}

Note that in the $\alpha=1$ (Skyrme-Faddeev) system, we can have vortices
without antivortex partners --- the restriction of the net vortex number being
zero does not apply in this case. Such Hopf-vortex fields have been the subject
of several studies, such as \cite{MNS00, HJS04, JH09}.
The results described above suggest that in the Skyrme-Faddeev system, the
normalized energy of a $p$-vortex decreases as $p$ increases. This indeed turns
out to be the case, as we now explain in more detail.

Consider fields $\vec\psi$ with energy density (\ref{Enden_Hopf})
on $\RR^2\times S^1$. So $\vec\psi$ is periodic in $x^3$ with period $L$, and
it satisfies the boundary condition $\vec\psi\to(0,0,1)$ as $\rho\to\infty$,
where $x^1+\ii x^2=\rho\ee^{\ii\theta}$.
More specifically, consider rotationally-symmetric $p$-vortices centred on
the $x^3$-axis: these will have the form
\begin{equation} \label{psi-vortex}
  \psi_1+\ii\psi_2=\sin{(f)}\,\exp(\ii p\theta+2\pi\ii x^3/L), \quad 
       \psi_3=\cos{(f)},
\end {equation}
where $f=f(\rho)$ satisfies $f(0)=\pi$ and $f(\rho)\to0$ as $\rho\to\infty$.
Finding the minimal energy $E(p)$ for various values of the vortex number $p$
is a straightforward numerical computation, and this was done for
$p=1,\ldots,5$. Let us, as usual, 
normalize the energy $E(p)$ by dividing it by the topological charge $p$.
The result is that $E(p)$ depends linearly on $1/p$:
\[
   E(p)\approx0.338/p+0.551;
\]
and it is this value (with, respectively, $p=1$ and $p=2$) which is used for
the $\alpha=1$ ends of the V+AV+V+AV and 2V+2AV plots in Figure~2.

In view of the above, it seems likely that triply-periodic solutions with even
lower values of $E_N$
than V+AV+V+AV and 2V+2AV could be constructed by assembling parallel arrays of
$p$-vortices and $p$-antivortices for $p\geq3$; but this has not been
investigated further.

The solutions corresponding to the final plot in Figure~2, referred to as
``multi-sheet'', have a lower value of energy-per-charge than the others
mentioned previously if $\alpha\geq0.9$, and do not resemble
parallel vortex-antivortex configurations for $\alpha$ close to 1. Figure~4 shows
the energy density isosurfaces where ${\cal E}$ equals 0.8 times its maximum value, for
four values of $\alpha$ ranging from 0.6 to 0.99.
\begin{figure}[hbt]
  \includegraphics[scale=0.45]{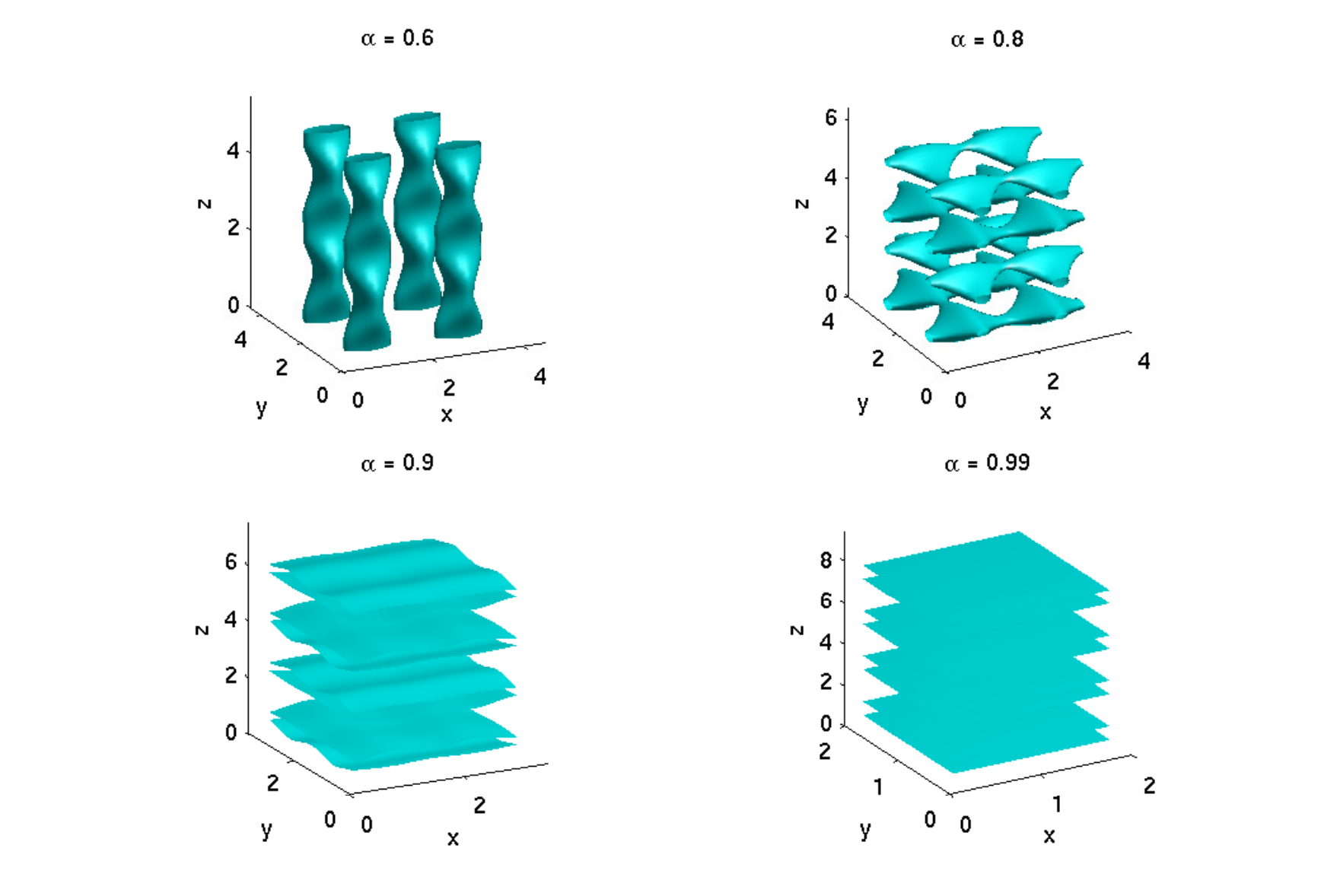}
  \caption{The ``multi-sheet" solution: energy densities for $\alpha=0.6,\, 0.8,\, 0.9,\, 0.99$.\label{fig4}}
\end{figure}
As $\alpha$ approaches 1, they are
closely approximated by fields which are homogeneous in $x^1$ and $x^2$, in fact
of the form
\begin{eqnarray}
Z_{1}:=\Phi_{1}+\ii\Phi_{2}&=&\sin(f) \exp(2\pi\ii\ve_2 x^2/L_2)\,, \label{newcrys1} \\
Z_{2}:=\Phi_{4}+\ii\Phi_{3}&=&\cos(f) \exp(2\pi\ii\ve_1 x^1/L_1)\,, \label{newcrys2} 
\end{eqnarray}
where $\ve_1=\pm1$, $\ve_2=\pm1$, and $f=f(x^3)$. Their energy density depends only
on $x^3$, and is peaked on sheets orthogonal to the $x^3$-axis.
Such fields arise if we impose the symmetries generated by
\begin{equation} \label{Sym2a}
x^1\mapsto x^1+c, \quad (Z_1,Z_2)\mapsto(Z_1,Z_2\exp(2\pi\ii\ve_1 c/L_1));
\end {equation}
\begin{equation} \label{Sym2b}
x^2\mapsto x^2+c, \quad (Z_1,Z_2)\mapsto (Z_1\exp(2\pi\ii\ve_2 c/L_2),Z_2);
\end {equation}
\begin{equation} \label{Sym2c}
(x^1,x^2)\mapsto(-x^1,-x^2), \quad (Z_1,Z_2)\mapsto(\overline{Z_1},\overline{Z_2}).
\end{equation}
Note that such transformations on  $(Z_1,Z_2)$ preserve the energy (\ref{Enden}).
In other words, translations in $x^1$ or $x^2$, and an inversion, can be
compensated by a symmetry transformation of the field.

In order to have periodicity in $x^3$, and for the topological charge $N$ to
be nonzero, we need to arrange things rather carefully. A periodic
field with $N=4$ is obtained by imposing $f(kL_3/4)=k\pi/2$ for $k=0,1,2,3,4$,
and
\begin{equation*}
  (\ve_1,\ve_2)=\left\{
   \begin{array}{ll}
       (1,1) & \mbox{for $0\leq x^3 < L_3/4$}\,, \\
       (-1,1) & \mbox{for $L_3/4\leq x^3 < L_3/2$}\,, \\
       (-1,-1) & \mbox{for $L_3/2\leq x^3 < 3L_3/4$}\,, \\
       (1,-1) & \mbox{for $3L_3/4\leq x^3 < L_3$}\,.
   \end{array}\right.
\end{equation*}
Note that the resulting field is continuous.

Substituting (\ref{newcrys1}, \ref{newcrys2}) into the formula (\ref{Enden})
for the energy density gives an expression of the form ${\cal E}=A(f)\,(f')^2+B(f)$,
where $f'=df/dx^3$. Then a standard Bogomolny argument implies that, for given
values of $L_1$ and $L_2$ (and $\alpha$), the normalized energy $E$ attains a
minimum value
\begin{equation} \label{Emin}
  E(\alpha,L_1,L_2) = 2 L_1 L_2 \int_0^{\pi/2}\sqrt{AB}\,df\,,
\end {equation}
when $L_3$ is given by
\begin{equation} \label{Lzmin}
  L_3 = 4\int_0^{\pi/2}\sqrt{A/B}\,df\,.
\end {equation}
The next step is to minimize (\ref{Emin}) with respect to $L_1$ and $L_2$,
which was done using numerical integration. The resulting minimal energy
for the ansatz (\ref{newcrys1}, \ref{newcrys2}) approaches that of the
multi-sheet solution (which is not quite homogeneous in $x^1$ and $x^2$)
as $\alpha\to1$. For example, if $\alpha=0.98$, then the ansatz energy
(\ref{Emin}) is only $0.6\%$ higher than that of the actual multi-sheet
solution, and the fields are almost identical.

For $\alpha\geq0.9$, the multi-sheet solution has the lowest
energy-per-charge of those that we investigated.
However, the $\alpha\to1$ limit of this solution, and of the
ansatz, is rather pathological, since the optimal values
of $L_1=L_2$ tend to zero. In fact, the energy (\ref{Emin}) equals
\begin{equation} 
  E = 2\pi\int_0^{\pi/2}\sqrt{2\lambda_2^2 \sin^2(2f) L_1^2
          + 4\pi^2 \lambda_2 \lambda_4 \sin^4(2f)}\,df
\end {equation}
when $\alpha=1$ (and where we have put $L_1=L_2$), from which it is clear that
its minimal value is attained when $L_1=0$. At first sight, this behaviour may
seem paradoxical, since the usual view is that the fourth-order Skyrme term in
the energy prevents an object with nontrivial 3-dimensional topology from
shrinking to zero volume. This view is indeed correct
for $0\leq\alpha<1$: in particular, if $\alpha$ lies in this range, then
a triply-periodic configuration with nonzero
topological charge $N$ has an energy which diverges if any of its periods tends
to zero. But in the $\alpha=1$ limit, namely for the Skyrme-Faddeev system,
this property no longer holds. One can see directly from
(\ref{newcrys1}, \ref{newcrys2}),
or rather its Hopf projection, why this is so. If one scales
each of the periods $L_j$, then each term in the expression of the energy of
$\vec\psi$ scales differently. In particular,
the term $\int(G_{12})^2$ has the form
\[
  \int(G_{12})^2\,dx^1\,dx^2\,dx^3 = K\frac{L_3}{L_1 L_2},
\]
where $K$ is its value when $L_1=L_2=L_3=1$. This is the only term which
prevents $L_1$ and $L_2$ from going to zero simultaneously. But for
(\ref{newcrys1}, \ref{newcrys2}) the quantity $G_{12}$ is identically zero, and
consequently one can always lower the energy by reducing $L_1$ and $L_2$.

\section{Conclusions}

We have studied triply-periodic stable solutions (local minima of the energy)
in a family of Skyrme-type systems parametrized by $\alpha\in[0,1]$, which
interpolates between the standard Skyrme model (at $\alpha=0$)  and the
Skyrme-Faddeev system (at $\alpha=1$).
At $\alpha=0$, the lowest-energy crystal resembles a cubic lattice of
half-skyrmion particles, and this picture persists near $\alpha=0$,
as one would expect. But for larger values of $\alpha$, various other
configurations are preferred (that is, have lower energy-per-charge):
for example, vortex-antivortex arrays and multi-sheet structures.

All these structures are somewhat problematic in the $\alpha\to1$ limit, and
it remains unclear whether or not there exists a smooth Skyrme-Faddeev
crystal which is at least stable under small perturbations. As remarked in the
introduction, the sublinear behaviour of the energy perhaps makes this unlikely.
But the possibility remains open, and is worth further investigation.


\bigskip\noindent{\bf Acknowledgments.}
Support from the UK Engineering and Physical Sciences Research Council
(under grant EP/G038775/1),
and the UK Science and Technology Facilities Council
(under ST/G000433/1), is gratefully acknowledged.


\end{document}